\documentclass{appolb}

\usepackage{graphicx}
\usepackage{subfigure}
\usepackage{bm}

\makeatletter
\newcommand\erfc{\mathop{\operator@font erfc}\nolimits}
\def\slashchar#1{\setbox0=\hbox{$#1$}
   \dimen0=\wd0 \setbox1=\hbox{/} \dimen1=\wd1
   \ifdim\dimen0>\dimen1 \rlap{\hbox to \dimen0{\hfil/\hfil}} #1
   \else  \rlap{\hbox to \dimen1{\hfil$#1$\hfil}} / \fi}

\makeatother

\begin{document}
\bibliographystyle{h-elsevier3} 
\title{Viscosity and dissipation - early stages\thanks{Talk presented at
the Workshop on Particle Correlations and Femtoscopy,
September  2008, Cracow, Poland. Supported by 
Polish Ministry of Science and Higher Education under
grant N202~034~32/0918.}}
\author{Piotr Bo\.zek
\footnote{email:~Piotr.Bozek@ifj.edu.pl}
\address{Institute of Physics, Rzesz\'ow University, 
PL-35959 Rzesz\'ow, Poland} 
\and
\address{Institute of Nuclear Physics PAN,
PL-31342 Krak\'ow, Poland} }
%

\maketitle

\begin{abstract}
 A very early start up time of the hydrodynamic evolution is needed in order 
to reproduce   observations
 from relativistic heavy-ion collisions experiments.
At such early times the systems is
 still not locally equilibrated. Another source of deviations from local 
equilibrium
is the viscosity of the fluid. We study these effects at very early times to 
obtain a dynamical prescription for the transition from an early 2-dimensional
 expansion to a nearly equilibrated 3-dimensional expansion at latter stages. 
The role of  viscosity at latter stages of the evolution 
is also illustrated.
\end{abstract}



\section{Introduction\label{sec:intro}}

Recent hydrodynamic calculations modelling heavy-ion collisions 
 can reproduce experimentally measured  soft observables~: 
transverse momentum spectra, collective elliptic flow and 
Hanbury-Brown Twiss correlation radii
 \cite{Chojnacki:2007rq,Broniowski:2008vp} if the initial 
time of the collective expansion is pushed down to $\tau_0=0.25$fm/c.
 This raises the question 
about the applicability of perfect the fluid hydrodynamics at such small proper 
times. The mechanism of the formation of the dense matter in the 
fireball is not  understood up to now. However, in
 all imaginable scenarios some 
time is required for the formation of the matter constituents and for their 
subsequent equilibration. In hydrodynamics, which is a coarse-grained 
description, the dynamics is defined by the local thermodynamical 
quantities, such as the  energy density and pressure. The details of the 
underlying microscopic degrees of freedom are irrelevant. Although formally,
 perfect fluid thermodynamics requires that local thermal equilibrium 
is maintained,  phenomenological applicability of the hydrodynamics 
in the description
 of heavy-ion collisions  starts as soon as the pressure becomes 
approximately isotropic. The dense matter in the fireball 
can be described by the hydrodynamic model after the time  when the
 effective pressure in the system is similar 
 in the longitudinal and transverse directions. Complete kinetic 
equilibrium is not required, since  the model has other sources 
limiting the robustness of its predictions, such as  the 
uncertainties in the high temperature equation of state, in the initial
density, and in the freeze-out procedure.

When the deviations of the energy momentum tensor $T^{\mu\nu}$
 from its form in a perfect fluid $T_0^{\mu\nu}$ 
\begin{equation}
T^{\mu\nu}=T_0^{\mu\nu}+\pi^{\mu\nu}
\end{equation}
is small the evolution
 can be formulated as the hydrodynamics of a viscous fluid 
\cite{Song:2007fn,Teaney:2003kp,Baier:2006sr,Baier:2006gy,Chaudhuri:2006jd,Muronga:2004sf}. But, in the very early evolution the initial anisotropy of the 
pressure is the main contribution that makes the matter to evolve differently 
from the perfect fluid  \cite{Bozek:2007di}. These early dissipative effects
 are strong, since the initial pressure anisotropy is large.

\section{Early dissipation}

The initial anisotropy of the pressure and its relaxation towards the perfect 
fluid value cannot be reliably described with the second order Israel-Stewart
relativistic viscous fluid formalism \cite {IS}. The applicability of the the
 viscous fluid equations requires $\pi^{\mu\nu}\pi_{\mu\nu} \ll p^2$, 
where $\pi^{\mu\nu}$ is the stress tensor. Instead, we propose an effective 
description of the transition from the anisotropic system with a two-dimensional 
pressure to the three-dimensional hydrodynamics \cite{Bozek:2007di}.
The energy momentum tensor is the sum of the perfect fluid 
energy momentum tensor and a stress correction
\begin{equation}
T^{\mu\nu}=\left( \begin{array}{cccc} \epsilon & 0 & 0 &0\\
0& p & 0 & 0\\
0& 0 & p & 0 \\
0 & 0 & 0 & p 
 \end{array}\right)+\left( \begin{array}{cccc} 0 & 0 & 0 &0\\
0& \pi/2 & 0 & 0\\
0& 0 & \pi/2 & 0 \\
0 & 0 & 0 & -\pi 
 \end{array}\right) \ .
\end{equation}
The dissipative correction $\pi$ quantifies the pressure anisotropy 
in the transverse and longitudinal directions. A similar form of the 
stress tensor appears in the hydrodynamics with shear viscosity for the
 case of the Bjorken flow \cite{Teaney:2003kp}. For large stress corrections 
the second order viscous hydrodynamics equations for $\pi$ cannot be reliably
 applied. Instead an
 effective equation describing the relaxation of 
the pressure asymmetry is used. Neglecting the shear viscosity we take
\begin{equation}
\pi(\tau)=\pi(\tau_0) e^{-(\tau-\tau_o)/\tau_\pi} ,
\label{eq:entropytpi}
\end{equation}
where $\tau_\pi$ is a phenomenological parameter, in principle unrelated to the 
relaxation time in the Isreal-Steward equation for the stress-tensor.

The dynamics is followed using a numerical solution of the relativistic 
hydrodynamic equations
\begin{equation}
\partial_\mu T^{\mu\nu}=0 
\end{equation}with some assumed symmetry of the fireball.
Entropy production from the dissipative relaxation of the pressure 
can be estimated in the Bjorken solution.
\begin{figure}
\includegraphics[width=.5\textwidth]{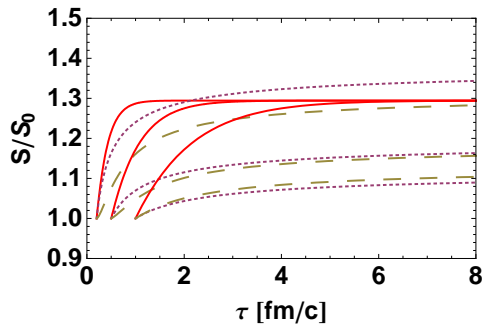}
\caption{Relative increase of the entropy from dissipative 
processes in the early stage of the collision 
for several initial times $\tau_0$ of the evolution. 
The dotted line represents the entropy production 
from the  Navier-Stokes shear viscosity tensor 
with $\eta=0.1\ s$, the  dashed 
line represents the increase of the entropy 
obtained from the second order viscous 
 hydrodynamic equation 
 with   $\eta=0.1 \ s$, $\tau_\pi=6 \eta/T s$,
 and $\Pi(\tau_0)=\frac{4\eta}{3\tau_0}$, and the solid 
represents the relative entropy production due to the 
stress tensor term of the form 
$\Pi(\tau)=p(\tau_0)\exp(-(\tau-\tau_0)/\tau_0)$ \cite{Bozek:2007di}.
\label{fig:entropy}}
\end{figure}
Depending on the ratio $\tau_\pi/\tau_0$, up to $30\%$ 
increase of the entropy is 
possible in the early phase. This additional entropy forces a retuning 
of the initial conditions of the evolution to reproduce final particle
 multiplicities. 
After this retuning is taken into account, most of the effect of the early
dissipation
 on final observables is canceled. However, we note that 
the transverse momentum 
spectra of final particles are harder if 
the early dissipative phase is present.
\begin{figure}
\includegraphics[width=.49\textwidth]{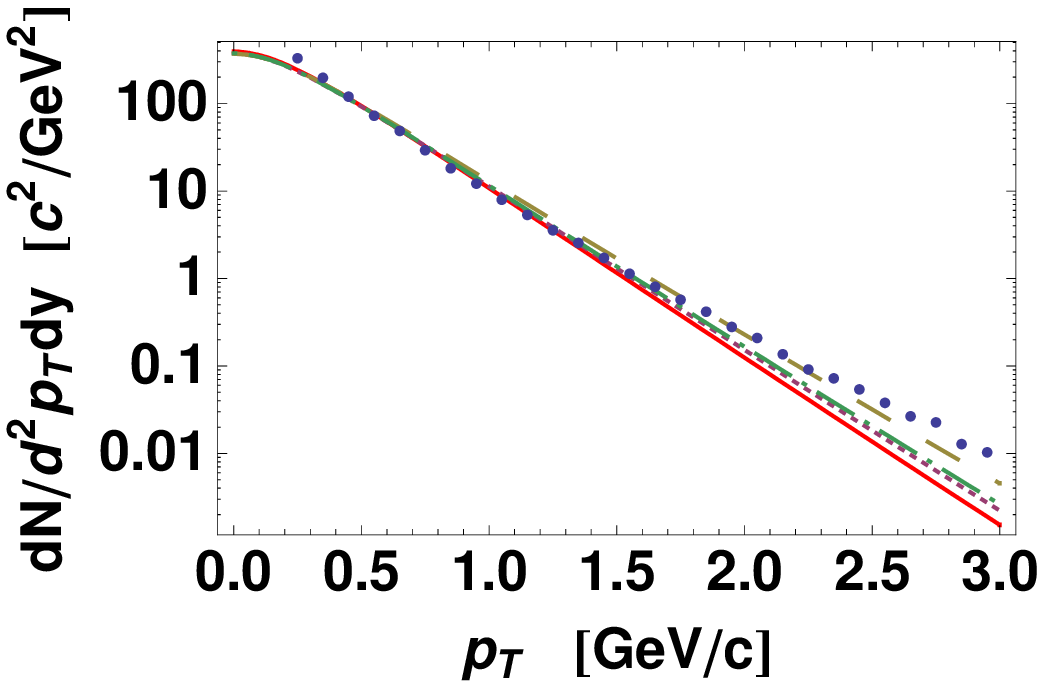}~\includegraphics[width=.49\textwidth]{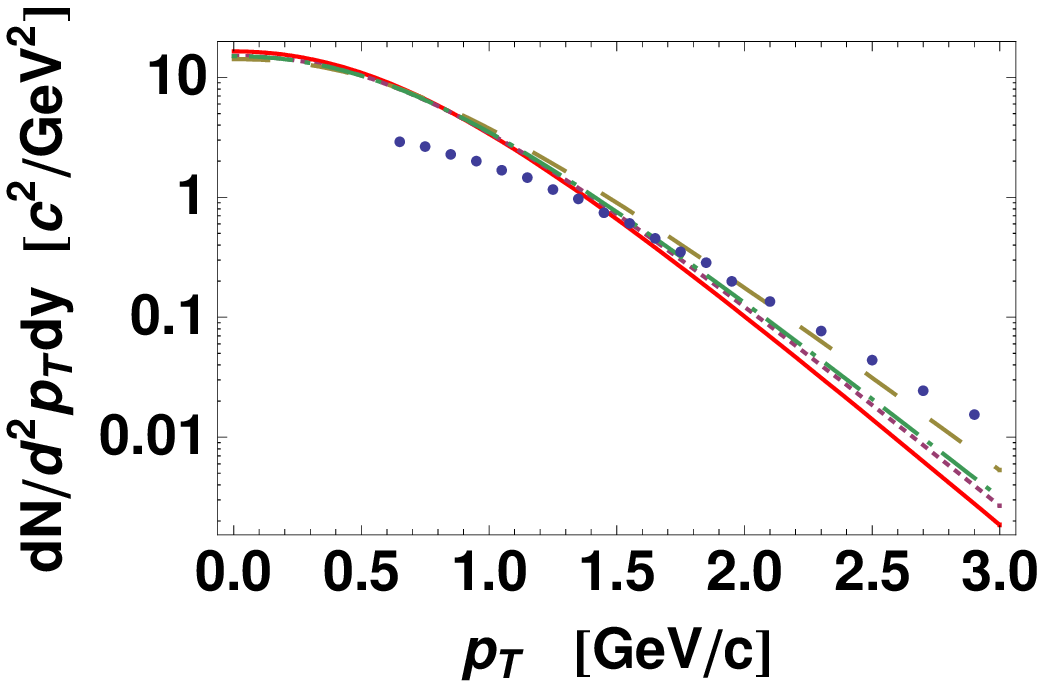}
\caption{$\pi^+$  (left) and proton (right) spectra 
from  hydrodynamic calculations 
(Solid and dashed-dotted line are for the ideal hydrodynamics
starting at $\tau_0=1$fm/c and $\tau_0=0.5$fm/c respectively.The 
dotted and dashed lines are  for the dissipative evolution corresponding to
$\tau_0=1$fm/c and $\tau_0=0.5$fm/c.). Data are from
 the PHENIX Collaboration \cite{Adler:2003cb} for most central events (0-5\%) \cite{Bozek:2007di}.
\label{fig:pip}}
\end{figure}

\section{Dissipation and viscosity}

We use relativistic hydrodynamics with viscosity  \cite{IS}. 
The stress tensor $\pi^{\mu\nu}$ is the solution of a dynamical equation
\begin{equation}
\tau_\pi \Delta^\mu_\alpha\Delta^\nu_\beta \pi^{\alpha\beta}+\pi^{\mu\nu}=
\eta < \nabla^\mu u^\nu >
-\frac{\eta T}{2\tau_\pi}\pi^{\mu\nu}\partial_\beta\left(\frac{\tau_\pi u^\beta}{\eta T}\right)
\end{equation}
where
\begin{equation}
< \nabla_\mu u_\nu >=
\nabla_\mu u_\nu+\nabla_\nu u_\mu-
\frac{2}{3}\Delta_{\mu\nu}\nabla_\alpha u^\alpha \ ,
\end{equation}
\begin{equation}
\nabla^\mu=\Delta^{\mu\nu}\partial_\nu
\end{equation}
with 
$u^\mu$ the fluid velocity, $\Delta_{\mu\nu}=g_{\mu\nu}-u_\mu u_\nu$, 
$\eta$ the shear viscosity, $\tau_\pi$ the relaxation time.
We solve the equations numerically in a boost-invariant geometry
 with an azimuthally asymmetric expansion in the transverse directions.
We use $\eta/s=1/4\pi$, $\tau_0=0.25$fm/c and
 $\pi^{zz}(\tau_0)/2=\pi^{xx}(\tau_0)=\pi^{yy}(\tau_0)=p/2$. 
Compared to other calculations of the hydrodynamic model with viscosity, 
we use a small initial time and a large value of the initial
 stress correction $\pi(\tau_0)$. 
\begin{figure}
\includegraphics[width=.6\textwidth]{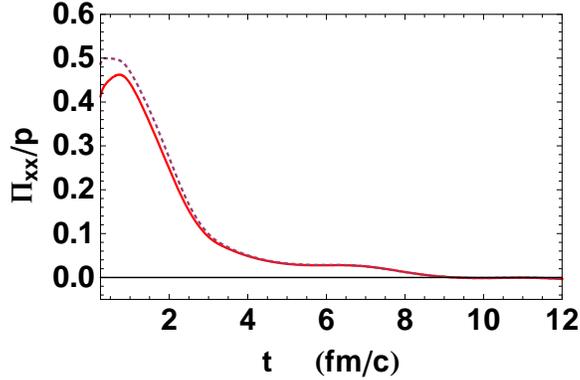}
\caption{Ratio of the stress correction to the pressure
 at the center of the fireball for two initial conditions for $\pi^{xx}$.}
\end{figure}
The model encompasses both the relaxation 
of the initial pressure anisotropy, and the latter interplay of the
 relaxation and velocity gradients. To compare with perfect fluid results
again  a retuning of the initial energy density is necessary to reproduce 
 the final multiplicity.
The additional transverse push is strong, it has a contribution from the
 initial stage of large pressure anisotropy and another one due to the
 viscosity driven stress corrections.
\begin{figure}
\includegraphics[width=.49\textwidth]{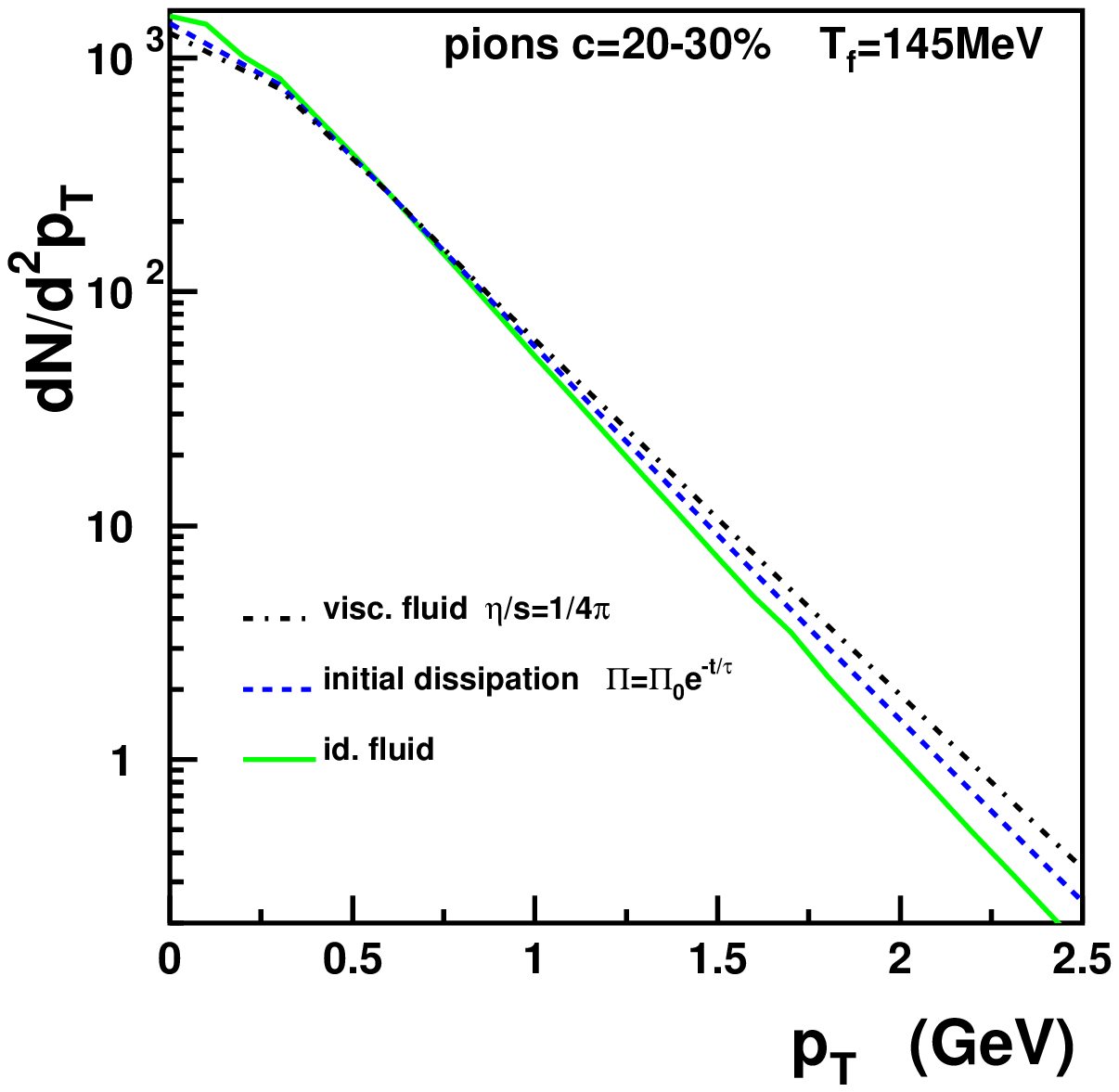}~\includegraphics[width=.49\textwidth]{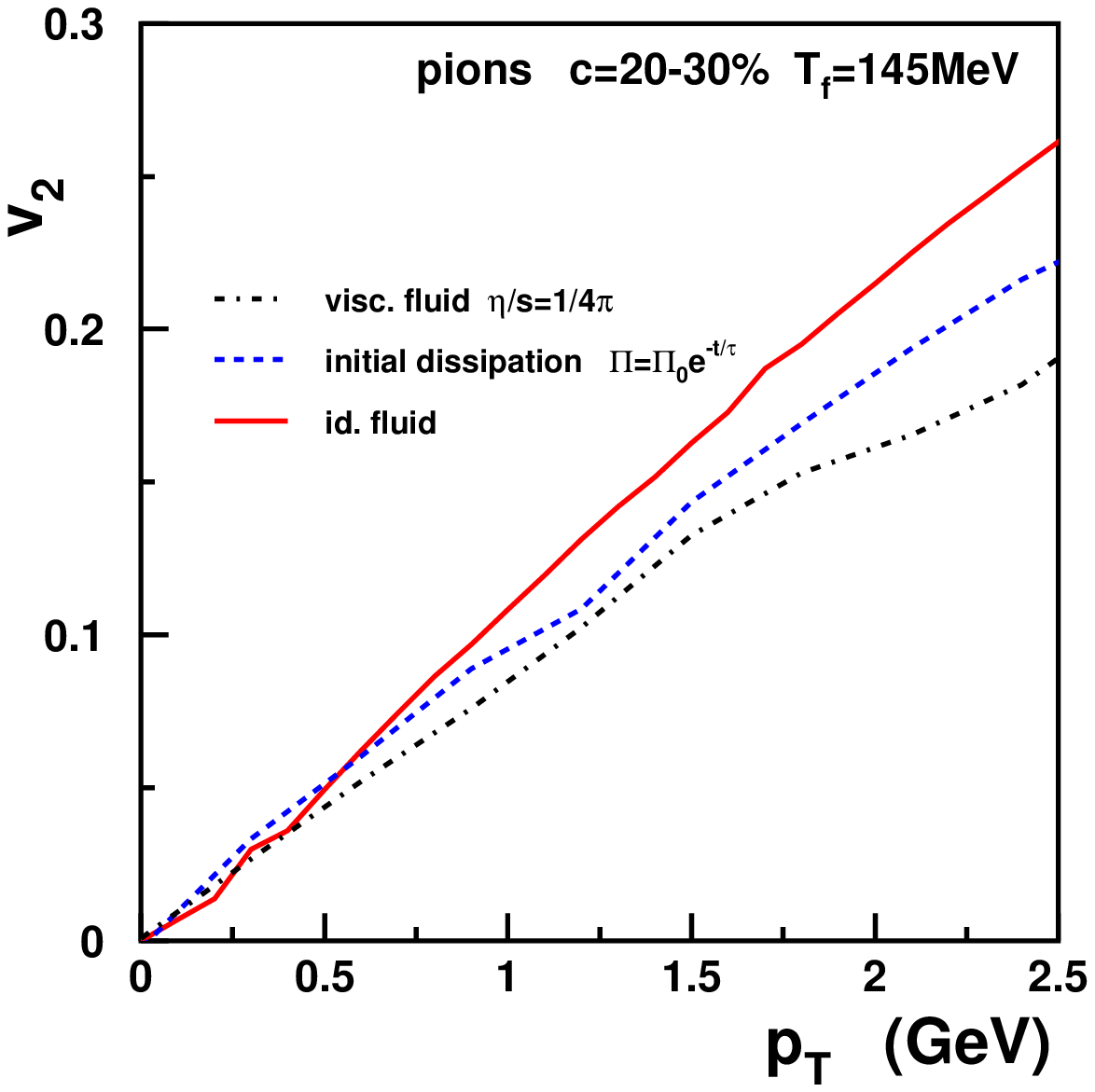}
\caption{Transverse momentum spectra (left) and elliptic flow coefficient (right)
 for $\pi^{+}$ for 
the perfect fluid (solid line), for the perfect fluid with initial
 pressure anisotropy (dashed line) and for the viscous fluid 
(dashed-dotted line).\label{fig:pt}}
\end{figure}
As a consequence of the prolongated transverse push, the transverse momentum 
spectra get even harder for the 
case when shear viscosity and initial anisotropy are 
combined 
than for the case with only initial dissipation. A similar effect
 is observed for the elliptic flow. The reduction of the azimuthal 
asymmetry is the strongest for the viscosity+dissipation scenario of the 
 fluid evolution.

\section{Summary}

We discuss dissipative effects in the very early phase of the collective 
development of the fireball created in relativistic heavy-ion collisions.
The initial anisotropy of the effective fluid pressure must dissipate. In the
 process entropy is produced. After the retuning of the initial 
conditions to accommodate for this additional entropy, the effect of the 
early dissipation
is most pronounced in the transverse momentum spectra of emitted particles.
The  initial dissipation of the pressure can be 
 taken together with the effect of 
the shear viscosity at latter stages. These corrections to the energy 
momentum-tensor combine to increase the transverse push in the collective 
flow and 
cause a significant reduction of the elliptic flow.

\bibliography{../hydr}

\end{document}